\documentclass[12pt, draftclsnofoot, onecolumn]{IEEEtran}
\usepackage{amssymb}
\usepackage{amsmath}
\usepackage{cite}
\usepackage{url}
\usepackage{xcolor}
\usepackage{cite,graphicx,amsmath,amssymb}
\usepackage{subfigure}
\usepackage{citesort}
\usepackage{fancyhdr}
\usepackage{mdwmath}
\usepackage{mdwtab}
\usepackage{caption}
\usepackage{amsthm}
\usepackage{setspace}
\usepackage{algorithm}
\usepackage{algorithmic}
\usepackage{makecell}
\usepackage{diagbox}
\newcommand{\bm}[1]{\mbox{\boldmath{$#1$}}}

\newtheorem{remark}{Remark}
\newtheorem{theorem}{Theorem}

\newtheorem{lemma}{Lemma}

\newtheorem{corollary}{Corollary}

\newtheorem{proposition}{Proposition}
\allowdisplaybreaks
\makeatletter
\def\ScaleIfNeeded{%
\ifdim\Gin@nat@width>\linewidth \linewidth \else \Gin@nat@width
\fi } \makeatother
\begin{document}
\title{Simultaneously Transmitting and Reflecting (STAR)-RISs: A Coupled Phase-Shift Model}
% \markboth{\textit{A Manuscript Submitted to The IEEE   Communications Magazine} }: Operating Protocols and Joint Beamforming Design
\vspace{-1cm}
\author{\IEEEauthorblockN{Yuanwei Liu\IEEEauthorrefmark{1}, Xidong Mu\IEEEauthorrefmark{2}, Robert Schober\IEEEauthorrefmark{3}, and H. Vincent Poor\IEEEauthorrefmark{4}}\\
\IEEEauthorblockA{\IEEEauthorrefmark{1}Queen Mary University of London, London, UK.\\ \IEEEauthorrefmark{2}Beijing University of Posts and Telecommunications, Beijing, China. \\\IEEEauthorrefmark{3}Friedrich-Alexander-University Erlangen-N{\"u}rnberg (FAU), Germany.\\ \IEEEauthorrefmark{4} Princeton University, Princeton, USA.\\E-mail: yuanwei.liu@qmul.ac.uk, muxidong@bupt.edu.cn, robert.schober@fau.de, poor@princeton.edu}}

\maketitle
\vspace{-1.5cm}
\begin{abstract}
A simultaneously transmitting and reflecting reconfigurable intelligent surface (STAR-RIS) aided communication system is investigated, where an access point sends information to two users located on each side of the STAR-RIS. Different from current works assuming that the phase-shift coefficients for transmission and reflection can be independently adjusted, which is non-trivial to realize for purely passive STAR-RISs, a coupled transmission and reflection phase-shift model is considered. Based on this model, a power consumption minimization problem is formulated for both non-orthogonal multiple access (NOMA) and orthogonal multiple access (OMA). In particular, the amplitude and phase-shift coefficients for transmission and reflection are jointly optimized, subject to the rate constraints of the users. To solve this non-convex problem, an efficient element-wise alternating optimization algorithm is developed to find a high-quality suboptimal solution, whose complexity scales only linearly with the number of STAR elements. Finally, numerical results are provided for both NOMA and OMA to validate the effectiveness of the proposed algorithm by comparing its performance with that of STAR-RISs using the independent phase-shift model and conventional reflecting/transmitting-only RISs.
\end{abstract}
%\begin{keywords}
%Coefficient design, coupled phase-shift model, reconfigurable intelligent surfaces, simultaneous transmission and reflection.
%\end{keywords}
%\footnotetext[1]{}
\section{Introduction}
Recently, the novel concept of simultaneously transmitting and reflecting reconfigurable intelligent surfaces (STAR-RISs)~\cite{STAR_mag} or intelligent omni-surfaces (IOSs)~\cite{IoS_mag} has been proposed\footnote{STAR-RISs and IOSs are based on a similar idea but have different hardware implementations. STAR-RISs are based on tunable metasurfaces~\cite{STAR_mag,tunable}, while IOSs employ positive intrinsic negative (PIN) diodes~\cite{IoS_mag}. In this paper, we focus on the category of metasurface-based STAR-RISs.}. In contrast to conventional reflecting-only RISs~\cite{RIS_survey}, the wireless signal incident on STAR-RISs is divided into the transmitted and reflected signals propagating into each side of the surface, thus achieving a \emph{full-space} reconfigurable wireless environment~\cite{STAR_mag}. Therefore, by deploying STAR-RISs, transmitters and receivers do not have to be located on the same side of the surface as is the case for conventional reflecting-only RISs, thus enhancing flexibility. Motivated by this promising characteristic, growing research efforts have been devoted to investigating the benefits of deploying STAR-RISs in wireless networks. For example, the authors of \cite{STAR} investigated the general hardware model and channel model for STAR-RISs, where the diversity gain achieved by STAR-RISs was analyzed. In \cite{protocol}, the authors proposed three practical operating protocols for STAR-RISs and studied the corresponding joint beamforming design problems in both unicast and multicast scenarios. Moreover, the fundamental coverage limits of STAR-RISs were characterized by the authors of \cite{chenyu}, where both non-orthogonal multiple access (NOMA) and orthogonal multiple access (OMA) schemes were considered.\\
\indent The key for achieving \emph{`STAR'} is that each element has to support both electric and magnetic currents to enable simultaneous transmission and reflection of the incident wireless signals~\cite{STAR_mag}. Moreover, the strengths and distributions of the two types of currents are determined by the electric and magnetic impedances of the STAR elements, whose realizable values mainly depend on the type of metasurface employed (e.g., passive, semi-passive, and active metasurfaces)~\cite{tunable}. All existing research contributions on STAR-RISs~\cite{STAR,protocol,chenyu} assume that the phase-shift coefficients for transmission and reflection can be \emph{independently} adjusted, which requires that the corresponding electric and magnetic impedances can assume arbitrary values. This, however, may be impossible for purely passive STAR-RISs whose realizable electric and magnetic impedances are limited to purely imaginary numbers~\cite{tunable}. In this case, the feasible phase-shift coefficients for transmission and reflection are \emph{coupled} with each other, which will not only cause performance degradation but also make the transmission and reflection coefficient/beamforming design much more complicated than in existing works~\cite{STAR,protocol,chenyu} which assume the independent phase-shift model. However, to the best of the authors' knowledge, none of the existing works characterizes the performance degradation caused by the coupled phase-shift model or investigates the corresponding transmission and reflection coefficient design. This is the motivation for this work.\\
\indent In this paper, we study a STAR-RIS aided downlink communication system, where an access point (AP) serves two users surrounding the STAR-RIS. We first introduce a coupled phase-shift model for purely passive and lossless STAR-RISs, where the absolute value of the phase-shift difference between transmission and reflection has to be either $\frac{1}{2}\pi $ or $\frac{3}{2}\pi$. Based on this model, we formulate a transmission and reflection coefficient design problem for NOMA and OMA, respectively, which aims to minimize the required power consumption for satisfying the rate requirements of the users. To solve this challenging problem, we propose an efficient element-wise alternating optimization (AO) algorithm, where the transmission/reflection phase-shift and amplitude coefficients of each STAR element are designed one by one to obtain a high-quality suboptimal solution. Numerical results comparing the performance of the proposed algorithm for the coupled phase-shift model with two benchmark schemes verify its effectiveness.\\
\indent \emph{Notations:} Scalars, vectors, and matrices are denoted by lower-case, bold-face lower-case, and bold-face upper-case letters, respectively. ${\mathbb{C}^{N \times 1}}$ denotes the space of $N \times 1$ complex-valued vectors. ${{\mathbf{a}}^*}$ and ${{\mathbf{a}}^H}$ denote the transpose and conjugate transpose of vector ${\mathbf{a}}$, respectively. ${\rm {diag}}\left( \mathbf{a} \right)$ denotes a diagonal matrix with the elements of vector ${\mathbf{a}}$ on the main diagonal. The distribution of a circularly symmetric complex Gaussian (CSCG) random variable with mean $\mu $ and variance ${\sigma ^2}$ is denoted by ${\mathcal{CN}}\left( {\mu,\sigma ^2} \right)$.
\section{System Model and Problem Formulation}%As shown in the right side of Fig. \ref{model},
\begin{figure}[!t]
  \centering
  \includegraphics[width=3.5in]{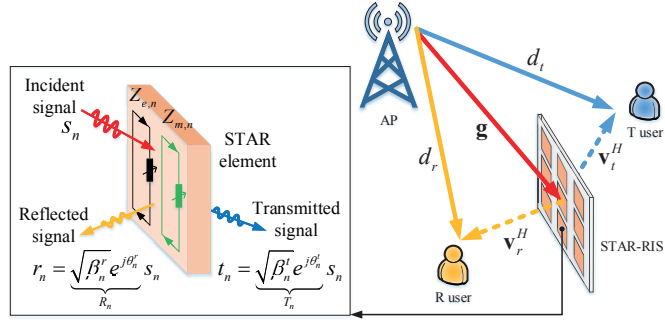}\\
  \caption{Illustration of a STAR-RIS aided downlink communication system.}\label{model}
\end{figure}

We consider a narrow-band STAR-RIS aided downlink communication system operating over frequency-flat channels, where a single-antenna AP transmits to two single-antenna users with the aid of a STAR-RIS comprising $N$ elements, see right hand side of Fig. \ref{model}. The left hand side (LHS) of Fig. \ref{model} depicts the architecture of a given STAR element. In particular, the STAR element is excited by the incident wireless signal, part of which is reflected into the same space as the incident signal, namely the reflected signal, and the other part of which is transmitted into the opposite space as the incident signal, namely the transmitted signal~\cite{STAR_mag}. Therefore, compared to conventional reflecting-only RISs, a full-space reconfigurable wireless environment can be facilitated by STAR-RISs. In this paper, we assume that one user is located in the transmission space of the STAR-RIS, referred to as T user, and the other is located at the reflection space of the STAR-RIS, referred to as R user. Let ${{d}}_k \in {{\mathbb{C}}^{1 \times 1}}$ and ${\mathbf{v}}_k \in {{\mathbb{C}}^{N \times 1}}$ denote the channels from the AP to user $k$ and the STAR-RIS to user $k$, respectively. Here, $k\in \left\{ {t,r} \right\}$ indicates the T and R users. Let ${\mathbf{g}} \in {{\mathbb{C}}^{N \times 1}}$ denote the channel between the AP and the STAR-RIS. To characterize the maximum performance gain of STAR-RISs, the perfect CSI of all channels is assumed to be available at the AP.
\vspace{-0.2cm}
\subsection{A Coupled Phase-shift Model for STAR-RISs}
Without loss of generality, let $s_n$ specify the signal incident on the $n$th STAR element, where $n \in {\mathcal{N}} \triangleq \left\{ {1,2, \ldots ,N} \right\}$. As illustrate on the LHS of Fig. \ref{model}, the corresponding transmitted and reflected signals from the $n$th STAR element are given by ${t_n}= {T_n}{s_n}$ and ${r_n}= {R_n}{s_n}$, respectively~\cite{STAR}. Here, ${T_n}= {\sqrt {\beta _n^t} {e^{j\theta _n^t}}}$ and ${R_n}= {\sqrt {\beta _n^r} {e^{j\theta _n^r}}}$ characterize the transmission and reflection coefficients of the $n$th STAR element, respectively, where $ {\beta _n^t}  \in \left[ {0,1} \right],\theta _n^t \in \left[ {0,2\pi } \right)$ and $ {\beta _n^r}  \in \left[ {0,1} \right],\theta _n^r \in \left[ {0,2\pi } \right)$ denote the amplitude and phase-shift adjustments imposed on the incident signal, $s_n$. Accordingly, the transmission- and reflection-coefficient matrices are given by ${{\mathbf{\Theta }}_t} = {\rm{diag}}\left( {{T_1},{T_2}, \ldots ,{T_N}} \right)$ and ${{\mathbf{\Theta }}_r} = {\rm{diag}}\left( {{R_1},{R_2}, \ldots ,{R_N}} \right)$, respectively\footnote{In this work, the time switching operating protocol~\cite{protocol} for STAR-RISs, where different transmission/reflection coefficients can be used in different time slots, is not considered. The reason for this is that frequently updating the transmission/reflection coefficients entails a high hardware complexity, which we want to avoid.}.\\
\indent To achieve the STAR function, each STAR element has to support both electric and magnetic currents~\cite{STAR_mag,tunable}, which can be characterized by one equivalent electric circuit and one equivalent magnetic circuit, as illustrated in Fig. \ref{model}. Let ${Z_{e,n}}$ and ${Z_{m,n}}$ denote the corresponding electric and magnetic impedances, respectively. Then, the transmission and reflection coefficients can be expressed in terms of these impedances as follows~\cite{tunable}:
\vspace{-0.2cm}
\begin{subequations}\label{TR}
\begin{align}\label{Tn}
&{T_n} = \frac{{2{Z_{e,n}}}}{{2{Z_{e,n}} + \eta }} - \frac{{{Z_{m,n}}}}{{{Z_{m,n}} + 2\eta }},  \\
\label{Rn}&{R_n} = \frac{{ - \eta }}{{2{Z_{e,n}} + \eta }} + \frac{{{Z_{m,n}}}}{{{Z_{m,n}} + 2\eta }},
\end{align}
\end{subequations}
\vspace{-1cm}

\noindent where $\eta$ denotes the free space wave impedance. By re-arranging \eqref{Tn} and \eqref{Rn}, we obtain~\cite{tunable}
\vspace{-0.2cm}
\begin{subequations}\label{ZME}
\begin{align}\label{Ze}
&{Z_{e,n}} = \frac{{\eta \left[ {1 + \left( {R_n + T_n} \right)} \right]}}{{2\left[ {1 - \left( {R_n + T_n} \right)} \right]}},  \\
\label{Zm}&{Z_{m,n}} = \frac{{2\eta \left[ {1 - \left( {R_n - T_n} \right)} \right]}}{{1 - \left( {R_n - T_n} \right)}},
\end{align}
\end{subequations}
\vspace{-1cm}

\noindent which specify the values of the electric and magnetic impedances needed to realize given transmission and reflection coefficients.\\
\indent Assuming that the STAR-RIS is a passive and lossless metasurface, the following two conditions have to be satisfied. First, ``lossless'' means that for each STAR element, the sum of the energies of the transmitted and reflected signals has to be equal to the energy of the incident signal, i.e., ${\left| {{t_n}} \right|^2} + {\left| {{r_n}} \right|^2} = {\left| {{s_n}} \right|^2}$, which implies the following condition for the transmission and reflection amplitude coefficients:
\vspace{-0.2cm}
\begin{align}\label{amplitude condition}
\beta _n^t + \beta _n^r = 1.
\end{align}
\vspace{-1cm}

\noindent This condition has also been considered in the existing works on STAR-RISs~\cite{STAR,protocol,chenyu}. Second, for a passive and lossless STAR-RIS, the electric and magnetic impedances are purely imaginary numbers~\cite{tunable}, i.e., $\operatorname{Re} \left( {{Z_{e,n}}} \right) = 0$ and $\operatorname{Re} \left( {{Z_{m,n}}} \right) = 0$. This constraint results in the following condition for both the amplitude and phase-shift coefficients\footnote{The derivation of \eqref{amplitude and phase condition} exploits \eqref{amplitude condition}. The details can be found in \cite{tunable}}:
\vspace{-0.2cm}
\begin{align}\label{amplitude and phase condition}
\sqrt {\beta _n^t} \sqrt {\beta _n^r} \cos \left( {\theta _n^t - \theta _n^r} \right) = 0,
\end{align}
\vspace{-1cm}

\noindent which indicates that the transmission and reflection phase-shift coefficients ($\theta _n^t$ and $\theta _n^r$) are coupled and have to satisfy the condition $\left| {\theta _n^t - \theta _n^r} \right| = \frac{\pi }{2}, \frac{{3\pi }}{2}$ if the real STAR mode is employed (i.e., $\beta _n^t \ne 0$ and $\beta _n^r \ne 0$). On the other hand, if the STAR element is operated in the pure transmission or reflection mode (i.e., $\beta _n^r = 0$ or $\beta _n^t = 0$), the remaining transmission or reflection phase-shift coefficients can be set to arbitrary values, which is consistent with conventional transmitting/reflecting-only RISs.
\begin{remark}\label{difference}
\emph{To enable independent adjustment of the transmission and reflection phase-shift coefficients in~\cite{STAR,protocol,chenyu}, the STAR-RISs have to be \emph{semi-passive} or \emph{active} to realize the required electric and magnetic impedances in \eqref{ZME}. Although the independent phase-shift model cannot be realized with purely passive STAR-RISs, it provides an important reference to evaluate the performance degradation caused by the coupled phase-shift model, see Section IV.}
\end{remark}
\subsection{System Model}
In this paper, we consider both NOMA and OMA. In NOMA, users share the same time/frequence resources and employ successive interference cancellation (SIC) to detect the superimposed signal delivered by the AP. In OMA, the users detect their intended signals independently by exploiting orthogonal time/frequence resources of equal size.
\subsubsection{NOMA} Let ${{{x}}_k}$, ${{{p}}_k}$, and ${c_k} \triangleq {d_k} + {\mathbf{v}}_k^H{{\mathbf{\Theta }}_k}{\mathbf{g}}$ denote the information-bearing symbol, the allocated power, and the effective channel for user $k \in \left\{ {t,r} \right\}$, respectively. The superimposed signal received at user $k\in \left\{ {t,r} \right\}$ is given by
\vspace{-0.2cm}
\begin{align}\label{received signal}
  {y_k}  = {c_k}\left( {{p_t}{x_t} + {p_r}{x_r}} \right) + {n_k},
\end{align}
\vspace{-1cm}

\noindent where ${n_k} \sim {\mathcal{CN}}\left( {0,\sigma ^2} \right)$ represents the additive white Gaussian noise (AWGN). In downlink NOMA, the user having the higher channel power gain (referred to as the strong user) first detects the signal of the user having the lower channel power gain (referred to as the weak user). Then, the strong user will subtract the detected weak user's signal from the received signal before detecting its own signal. Therefore, the achievable communication rate of user $k\in \left\{ {t,r} \right\}$ is given by
\vspace{-0.2cm}
\begin{align}\label{rate NOMA}
R_k^{\rm {N}} = {\log _2}\left( {1 + \frac{{{{\left| {{c_k}} \right|}^2}{p_k}}}{{{\lambda _k}{{\left| {{c_k}} \right|}^2}{p_{\overline k }} + {\sigma ^2}}}} \right),
\end{align}
\vspace{-0.8cm}

\noindent where $\overline k  = r$, if $k = t$; and $\overline k  = t$, otherwise. Here, binary variables, ${\lambda _k} \in \left\{ {0,1} \right\}, k\in \left\{ {t,r} \right\}$, indicate the decoding order of the two users and ${\lambda _t} + {\lambda _s}=1$. If ${\left| {{c_t}} \right|^2} \ge {\left| {{c_r}} \right|^2}$ (i.e., T user is the strong user), we have ${\lambda _t} = 0$ and ${\lambda _r} = 1$; otherwise, ${\lambda _t} = 1$ and ${\lambda _r} = 0$.
\subsubsection{OMA} By employing orthogonal time/frequence resources, OMA avoids inter-user interference. Therefore, the corresponding achievable communication rate of user $k\in \left\{ {t,r} \right\}$ is given by
\vspace{-0.2cm}
\begin{align}\label{rate OMA}
R_k^{\rm {O}} = \frac{1}{2}{\log _2}\left( {1 + \frac{{{{\left| {{c_k}} \right|}^2}{p_k}}}{{\frac{1}{2}{\sigma ^2}}}} \right).
\end{align}
\subsection{Problem Formulation}
Our goal is to minimize the total power consumption of the AP by jointly optimizing the transmission and reflection coefficients, subject to the rate requirements of both users and the coupled constraints on the transmission and reflection coefficients. Therefore, the optimization problem can be formulated as follows:
\vspace{-0.2cm}
\begin{subequations}\label{P1 X}
\begin{align}
&\mathop {\min }\limits_{ {{{{p}}_k},{\mathbf{\Theta }}_k, {\lambda _k}} } \;\;{p_t} + {p_r}  \\
\label{QoS X}{\rm{s.t.}}\;\;& R_k^{\rm{X}} \ge {\overline R _k},\forall k \in \left\{ {t,r} \right\},\\
\label{amplitude}&\beta _n^t + \beta _n^r = 1,\forall n \in {\mathcal{N}},\\
\label{phase}&\left| {\theta _n^t - \theta _n^r} \right| = \frac{1}{2}\pi \;{\rm{or}}\;\frac{3}{2}\pi ,\forall n \in {\mathcal{N}},\\
\label{range}&\theta _n^t,\theta _n^r \in \left[ {0,2\pi } \right), {\beta _n^t} , {\beta _n^r}  \in \left[ {0,1} \right],\forall n \in {\mathcal{N}},\\
\label{order1}&\left\{ \begin{gathered}
  {\lambda _t} = 0,{\lambda _r} = 1,\;{\rm{if}}\;\;{\left| {{c_t}} \right|^2} \ge {\left| {{c_r}} \right|^2}, \hfill \\
  {\lambda _t} = 1,{\lambda _r} = 0,\;{\rm{otherwise}}, \hfill \\
\end{gathered}  \right.\\
\label{order2}&{\lambda _t},{\lambda _r} \in \left\{ {0,1} \right\},{\lambda _t} + {\lambda _r} = 1,
\end{align}
\end{subequations}
\vspace{-1cm}

\noindent where ${\overline R _k}$ denotes the minimum rate requirement of user $k$ and ${\rm{X}} \in \left\{ {{\rm{N}},{\rm{O}}} \right\}$ indicates whether NOMA or OMA is employed. Constraints \eqref{amplitude} and \eqref{phase} are the coupled amplitude and phase-shift coefficients, respectively. Variable ${\lambda _k}$ and constraints \eqref{order1} and \eqref{order2} are only valid when NOMA is employed. Note that for problem \eqref{P1 X}, we assume that the coupled phase-shift constraint \eqref{phase} always exists even if ${\beta _n^t}=0$ or ${\beta _n^r}=0$, while in fact in this case, $\theta _n^t$ and $\theta _n^r$ can be independently adjusted according to \eqref{amplitude and phase condition}. However, \eqref{phase} will not lead to any performance degradation compared to \eqref{amplitude and phase condition}. To elaborate this, let us take ${\beta _n^t}=1$ and ${\beta _n^r}=0$ as an example. In this case, ${\beta _n^r}=0$ causes $\theta _n^r$ to become a dummy variable and only $\theta _n^t$ has an impact on the solution of \eqref{P1 X}. In particular, in this case, $\theta _n^t$ can assume any arbitrary value from $\left[ {0,2\pi } \right)$, since the dummy variable, $\theta _n^r$, can be set accordingly to satisfy \eqref{phase} without affecting the solution. For the case of ${\beta _n^t}=0$ and ${\beta _n^r}=1$, a similar conclusion can be drawn. Therefore, in this special case, \eqref{phase} will not influence the achievable performance and \eqref{P1 X} is equivalent to the corresponding problem employing the independent phase-shift model.
\begin{remark}\label{difference 2}
\emph{The main challenges for solving problem \eqref{P1 X} can be summarized as follows. On the one hand, compared to conventional reflecting-only RISs~\cite{Zheng_NOMA}, where only the reflection phase-shift coefficients have to be optimized, STAR-RISs require the joint optimization of both amplitude and phase-shift coefficients for transmission and reflection. On the other hand, compared to STAR-RISs employing the independent phase-shift model~\cite{protocol}, the coupled phase-shift constraint \eqref{phase} further complicates the design of the transmission and reflection coefficients. Therefore, the algorithms proposed in the existing works~\cite{Zheng_NOMA,protocol} cannot be applied for solving \eqref{P1 X}.}
\end{remark}
\section{Proposed Solution}
To solve problem \eqref{P1 X}, we first transform it into a tractable form. Note that for the optimal solution of \eqref{P1 X}, the inequality rate constraint \eqref{QoS X} has to be satisfied with equality. Therefore, problem \eqref{P1 X} can be rewritten as follows:
\vspace{-0.3cm}
\begin{subequations}\label{P2-X}
\begin{align}
&\mathop {\min }\limits_{ {{\mathbf{\Theta }}_k, {\lambda _k}} } \;\;{P^{\rm{X}}}  \\
\label{C P2-X}{\rm{s.t.}}\;\;& \eqref{amplitude}-\eqref{order2},
\end{align}
\end{subequations}
\vspace{-1cm}

\noindent where ${P^{\rm{X}}}$ for NOMA and OMA is respectively given by
\vspace{-0.3cm}
\begin{subequations}\label{Power}
\begin{align}\label{Power NOMA}
{P^{\rm{N}}}& = \left\{ \begin{gathered}
  \frac{{{{\overline \gamma _r^{\rm{N}}  }}}}{{{{\left| {{c_r}} \right|}^2}}} + \frac{{{{\overline \gamma _t^{\rm{N}}  }}\left( {{{\overline \gamma _r^{\rm{N}}  }} + 1} \right)}}{{{{\left| {{c_t}} \right|}^2}}} \triangleq P_{{\lambda _t} = 0,{\lambda _r} = 1}^{\rm{N}}, \hfill \\
  \frac{{{{\overline \gamma _t^{\rm{N}}  }}}}{{{{\left| {{c_t}} \right|}^2}}} + \frac{{{{\overline \gamma _r^{\rm{N}}  }}\left( {{{\overline \gamma _t^{\rm{N}}  }} + 1} \right)}}{{{{\left| {{c_r}} \right|}^2}}}\triangleq P_{{\lambda _t} = 1,{\lambda _r} = 0}^{\rm{N}}, \hfill \\
\end{gathered}  \right.\\
\label{Power OMA}{P^{\rm{O}}}& = \frac{{\overline \gamma  _t^{\rm{O}}}}{{{{\left| {{c_t}} \right|}^2}}} + \frac{{\overline \gamma  _r^{\rm{O}}}}{{{{\left| {{c_r}} \right|}^2}}},
\end{align}
\end{subequations}
\vspace{-1cm}

\noindent where $\overline \gamma  _k^{\rm{N}} \triangleq \left( {{2^{{{\overline R }_k}}} - 1} \right){\sigma ^2}$ and $\overline \gamma  _t^{\rm{O}} \triangleq {{\left( {{2^{2{{\overline R }_k}}} - 1} \right){\sigma ^2}} \mathord{\left/
 {\vphantom {{\left( {{2^{2{{\overline R }_k}}} - 1} \right){\sigma ^2}} 2}} \right.
 \kern-\nulldelimiterspace} 2}$, $\forall k \in \left\{ {t,r} \right\}$. As can be observed, objective functions for NOMA and OMA share a similar structure while NOMA requires additional decoding order constraints \eqref{order1} and \eqref{order2}. As a result, in the following, we mainly focus on solving problem \eqref{P2-X} for NOMA since the proposed algorithm is also applicable to OMA. Recall that there are $2! = 2$ possible decoding orders for problem \eqref{P2-X} for NOMA.  The optimal solution can be obtained by exhaustively searching over the two options, i.e., ${P^{{\rm{N}}}} = \min \left( {P_{{\lambda _t} = 0,{\lambda _r} = 1}^{{\rm{N}}},P_{{\lambda _t} = 1,{\lambda _r} = 0}^{{\rm{N}}}} \right)$. Before solving \eqref{P2-X}, we first introduce the following proposition.
\begin{proposition}\label{AB}
\emph{For the optimal solution of problem \eqref{P2-X}, the NOMA decoding order constraints \eqref{order1} and \eqref{order2} are automatically satisfied.}
\begin{proof}
\emph{Let $\left\{ {{\mathbf{\Theta }}_k^{op},k \in \left\{ {t,r} \right\}} \right\}$ denote the optimal solution of problem \eqref{P2-X}. Then, we have}
\begin{align}
\begin{gathered}
  P_{{\lambda _t} = 0,{\lambda _r} = 1}^{\rm{N}}\left( {\left\{ {{\mathbf{\Theta }}_k^{op}} \right\}} \right) - P_{{\lambda _t} = 1,{\lambda _r} = 0}^{\rm{N}}\left( {\left\{ {{\mathbf{\Theta }}_k^{op}} \right\}} \right) \hfill \\
   = {\overline \gamma  _t}{\overline \gamma  _r}\left( {\frac{1}{{{{\left| {{c_t}\left( {\left\{ {{\mathbf{\Theta }}_k^{op}} \right\}} \right)} \right|}^2}}} - \frac{1}{{{{\left| {{c_r}\left( {\left\{ {{\mathbf{\Theta }}_k^{op}} \right\}} \right)} \right|}^2}}}} \right)\triangleq \delta,  \hfill \\
\end{gathered}
\end{align}
\emph{where ${{{\left| {{c_t}\left( {\left\{ {{\mathbf{\Theta }}_k^{op}} \right\}} \right)} \right|}^2}}$ and ${{{\left| {{c_r}\left( {\left\{ {{\mathbf{\Theta }}_k^{op}} \right\}} \right)} \right|}^2}}$ denote the corresponding effective channel power gain for the T and R users, respectively. If ${{\lambda _t} = 0,{\lambda _r} = 1}$ is the optimal decoding order, ${\left| {{c_t}\left( {\left\{ {{\mathbf{\Theta }}_k^{op}} \right\}} \right)} \right|^2} \ge {\left| {{c_r}\left( {\left\{ {{\mathbf{\Theta }}_k^{op}} \right\}} \right)} \right|^2}$ has to hold such that $\delta  \le 0$ (i.e., $P_{{\lambda _t} = 0,{\lambda _r} = 1}^{\rm{N}}\left( {\left\{ {{\mathbf{\Theta }}_k^{op}} \right\}} \right) \le P_{{\lambda _t} = 1,{\lambda _r} = 0}^{\rm{N}}\left( {\left\{ {{\mathbf{\Theta }}_k^{op}} \right\}} \right)$). Otherwise, ${{\lambda _t} = 0,{\lambda _r} = 1}$ cannot be the optimal decoding order. The case of ${{\lambda _t} = 1,{\lambda _r} = 0}$ can be proved in a similar manner.}
\end{proof}
\end{proposition}
By exploiting \textbf{Proposition \ref{AB}}, we can remove constraints \eqref{order1} and \eqref{order2} from problem \eqref{P2-X}. In the following, for simplicity of presentation, we consider the case ${\lambda _t} = 0,{\lambda _r} = 1$ as an example. The resulting optimization problem simplifies to\footnote{For NOMA with ${\lambda _t} = 1,{\lambda _r} = 0$ and OMA, we just need to replace the objective function of \eqref{P3-X} with $P_{{\lambda _t} = 1,{\lambda _r} = 0}^{\rm{N}}$ and $P^{\rm{O}}$ from \eqref{Power}.}
\vspace{-0.3cm}
\begin{subequations}\label{P3-X}
\begin{align}
&\mathop {\min }\limits_{ {{\mathbf{\Theta }}_k}} \;\;\frac{{{{\overline \gamma _r^{\rm{N}}}}}}{{{{\left| {{c_r}} \right|}^2}}} + \frac{{{{\overline \gamma _t^{\rm{N}}}}\left( {{{\overline \gamma _r^{\rm{N}}}} + 1} \right)}}{{{{\left| {{c_t}} \right|}^2}}} \\
\label{C P3-X1}{\rm{s.t.}}\;\;& \eqref{amplitude}-\eqref{range}.
\end{align}
\end{subequations}
\vspace{-1cm}

\noindent However, it is still challenging to solve problem \eqref{P3-X}, since not only the objective function is non-convex with respect to ${{\mathbf{\Theta }}_k}$ but also the feasible set of the transmission and reflection coefficients. Since it is difficult to find a globally optimal solution for such a challenging problem, in the following, we decompose \eqref{P3-X} into two subproblems, namely a phase-shift coefficient design problem and a amplitude coefficient design problem. For each subproblem, an element-wise AO algorithm will be developed to find a high-quality suboptimal solution.
\subsubsection{Phase-shift Coefficient Design} To start with, we first define transmission/reflection amplitude and phase-shift vectors ${{\bm{\beta}} _k} \triangleq \left[ {\sqrt{\beta _1^k},\sqrt{\beta _2^k}, \ldots ,\sqrt{\beta _N^k}} \right]^T$ and ${{\mathbf{q}}_k} = {\left[ {{e^{j\theta _1^k}},{e^{j\theta _2^k}}, \ldots ,{e^{j\theta _N^k}}} \right]^T},$\\$\forall k \in \left\{ {t,r} \right\}$, respectively. Therefore, the transmission/reflection-coefficient matrix can be re-expressed as ${{\mathbf{\Theta}} _k} = {\rm{diag}}\left( {{{\bm{\beta}} _k}} \right){\rm{diag}}\left( {{{\mathbf{q}}_k}} \right)$. Accordingly, the effective channel power gain of user $k \in \left\{ {t,r} \right\}$ can be rewritten as follows:
\vspace{-0.2cm}
\begin{align}\label{received signal2}
  {\left| {{c_k}} \right|^2}  = {\left| {{d_k} + {\mathbf{v}}_k^H{\rm{diag}}\left( {{{\bm{\beta}} _k}} \right){\rm{diag}}\left( {{{\mathbf{q}}_k}} \right){\mathbf{g}}} \right|^2}  = {\left| {{\mathbf{s}}_k^H{{\overline {\mathbf{q}} }_k}} \right|^2},
\end{align}
\vspace{-1cm}

\noindent where ${\mathbf{s}}_k^H \triangleq \left[ {{\mathbf{v}}_k^H{\rm{diag}}\left( {{{\bm{\beta}} _k}} \right){\rm{diag}}\left( {\mathbf{g}} \right)\;{d_k}} \right] \in {{\mathbb{C}}^{1 \times \left( {N + 1} \right)}}$ and ${\overline {\mathbf{q}} _k} \triangleq {\left[ {{\mathbf{q}}_k^T\;1} \right]^T}\in {{\mathbb{C}}^{\left( {N + 1} \right) \times 1}}$. For given ${{\bm{\beta}} _k}$, problem \eqref{P3-X} reduces to the following transmission and reflection phase-shift coefficient design problem:
\vspace{-0.3cm}
\begin{subequations}\label{P4-X}
\begin{align}
&\mathop {\min }\limits_{{{\overline {\mathbf{q}} }_k},{\psi _k}} \;\; \frac{{\overline \gamma _r^{\rm{N}}}}{{{\psi _r}}} + \frac{{\overline \gamma _t^{\rm{N}}\left( {\overline \gamma _r^{\rm{N}} + 1} \right)}}{{{\psi _t}}}\\
\label{C P4-X1}{\rm{s.t.}}\;\;&{\left| {{\mathbf{s}}_k^H{{\overline {\mathbf{q}} }_k}} \right|^2} \ge {\psi _k},\forall k \in \left\{ {t,r} \right\},\\
\label{C P4-X2}&{\left[ {{{\mathbf{q}}_t}} \right]_n} = j{\left[ {{{\mathbf{q}}_r}} \right]_n}\;{\rm{or}}\;{\left[ {{{\mathbf{q}}_t}} \right]_n} =  - j{\left[ {{{\mathbf{q}}_r}} \right]_n},\forall n \in {{\mathcal{N}}}, \\
\label{C P4-X3}&\left| {{{\left[ {{{\mathbf{q}}_t}} \right]}_n}} \right| = 1,\left| {{{\left[ {{{\mathbf{q}}_r}} \right]}_n}} \right| = 1,\forall n \in {{\mathcal{N}}},
\end{align}
\end{subequations}
\vspace{-1cm}

\noindent where ${\left\{ {{\psi _k}} \right\}}$ are auxiliary variables, ${j^2} =  - 1$, and ${\left[  \cdot  \right]_n}$ represents the $n$th element of a vector. Here, constraint \eqref{C P4-X2} is equivalent to constraint \eqref{phase}, where ${\left[ {{{\mathbf{q}}_t}} \right]_n} = j{\left[ {{{\mathbf{q}}_r}} \right]_n}$ implies that $\left| {\theta _n^t - \theta _n^r} \right| = \frac{1}{2}\pi $ and ${\left[ {{{\mathbf{q}}_t}} \right]_n} =  - j{\left[ {{{\mathbf{q}}_r}} \right]_n}$ implies that $\left| {\theta _n^t - \theta _n^r} \right| = \frac{3}{2}\pi $. Although the objective function of \eqref{P4-X} is convex, it is still a non-convex problem since the LHS of \eqref{C P4-X1} is not concave and constrains \eqref{C P4-X2} and \eqref{C P4-X3} are non-convex. To handle this obstacle, we propose an element-wise AO algorithm. For any one of the elements in ${{\mathbf{q}}_k}$ (i.e., $\left[ {{{\mathbf{q}}_k}} \right]_n$) with the other $N-1$ elements (i.e., $\left\{ {{{\left[ {{{\mathbf{q}}_k}} \right]}_l}} \right\}_{l \ne n}^N$) fixed, ${\left| {{\mathbf{s}}_k^H{{\overline {\mathbf{q}} }_k}} \right|^2}$ can be expressed as follows:
\begin{align}
  {\left| {{\mathbf{s}}_k^H{{\overline {\mathbf{q}} }_k}} \right|^2} = {A_{k,n}} + 2\operatorname{Re} \left\{ {{B_{k,n}}{{\left[ {{{\mathbf{q}}_k}} \right]}_n}} \right\},
\end{align}
where ${A_{k,n}} \triangleq {\left| {\sum\nolimits_{l \ne n}^{N + 1} {{{\left[ {{\mathbf{s}}_k^H} \right]}_l}} {{\left[ {{{\mathbf{q}}_k}} \right]}_l}} \right|^2} + {\left| {{{\left[ {{\mathbf{s}}_k^H} \right]}_n}} \right|^2}$, ${B_{k,n}} \triangleq {\left[ {{\mathbf{s}}_k^H} \right]_n}{\left\{ {\sum\nolimits_{l \ne n}^{N + 1} {{{\left[ {{\mathbf{s}}_k^H} \right]}_l}} {{\left[ {{{\mathbf{q}}_k}} \right]}_l}} \right\}^*}$, and ${\left[ {{{\mathbf{q}}_k}} \right]_{N + 1}} \triangleq 1, \forall k \in \left\{ {t,r} \right\},n \in {{\mathcal{N}}}$. In this case, ${\left| {{\mathbf{s}}_k^H{{\overline {\mathbf{q}} }_k}} \right|^2}$ is an affine function with respect to $\left[ {{{\mathbf{q}}_k}} \right]_n$. For given $N-1$ transmission and reflection phase-shift coefficients, the design of the $n$th transmission and reflection phase-shift coefficients can be formulated as follows:
\begin{subequations}\label{P5-X}
\begin{align}
&\mathop {\min }\limits_{{{\left[ {{{\mathbf{q}}_r}} \right]}_n},{\psi _k}} \;\; \frac{{\overline \gamma _r^{\rm{N}}}}{{{\psi _r}}} + \frac{{\overline \gamma _t^{\rm{N}}\left( {\overline \gamma _r^{\rm{N}} + 1} \right)}}{{{\psi _t}}}\\
\label{C P5-X1}{\rm{s.t.}}\;\;&{A_{r,n}} + 2\operatorname{Re} \left\{ {{B_{r,n}}{{\left[ {{{\mathbf{q}}_r}} \right]}_n}} \right\} \ge {\psi _r},\\
\label{C P5-X2}&{A_{t,n}} + 2\operatorname{Re} \left\{ {{B_{t,n}}\left\{ { \pm j{{\left[ {{{\mathbf{q}}_r}} \right]}_n}} \right\}} \right\} \ge {\psi _t}, \\
\label{C P5-X3}&\left| {{{\left[ {{{\mathbf{q}}_r}} \right]}_n}} \right| = 1.
\end{align}
\end{subequations}
As there are two possible phase-shift differences between ${{{\left[ {{{\mathbf{q}}_t}} \right]}_n}}$ and ${{{\left[ {{{\mathbf{q}}_r}} \right]}_n}}$, we can first solve the two subproblems and then select the optimal solution from the two results. For each subproblem, the non-convexity only lies in the unit-modulus constraint \eqref{C P5-X3}. To address this issue, we can relax the non-convex unit-modulus constraint \eqref{C P5-X3} into a convex one, i.e., $\left| {{{\left[ {{{\mathbf{q}}_r}} \right]}_n}} \right| \le 1$. By exploiting the result in \cite[Appendix B]{shuowen}, it can be shown that the optimal solution to the relaxed problem will always satisfy $\left| {{{\left[ {{{\mathbf{q}}_r}} \right]}_n}} \right| = 1$. The relaxed problem is a standard convex optimization problem, which can be efficiently solved via standard convex problem solvers such as CVX~\cite{cvx}. By alternatingly optimizing each phase-shift coefficient ${\left[ {{{\mathbf{q}}_r}} \right]_n}$ with all the other $\left\{ {{{\left[ {{{\mathbf{q}}_k}} \right]}_l}} \right\}_{l \ne n}^N$ fixed, a suboptimal solution to \eqref{P4-X} can be obtained.
\subsubsection{Amplitude Coefficient Design} Similarly, for optimizing the transmission and reflection amplitude coefficients, we first rewrite the effective channel power gain of user $k \in \left\{ {t,r} \right\}$ as follows:
\begin{align}\label{channel power gain amplitude}
{\left| {{c_k}} \right|^2}  = {\left| {{d_k} + {\mathbf{v}}_k^H{\rm{diag}}\left( {{{\bm{\beta}} _k}} \right){\rm{diag}}\left( {{{\mathbf{q}}_k}} \right){\mathbf{g}}} \right|^2}  = {\left| {{\mathbf{b}}_k^H{{\overline {\bm{\beta}} }_k}} \right|^2},
\end{align}
where ${\mathbf{b}}_k^H \triangleq \left[ {{\mathbf{v}}_k^H{\rm{diag}}\left( {{{\mathbf{q}} _k}} \right){\rm{diag}}\left( {\mathbf{g}} \right)\;{d_k}} \right] \in {{\mathbb{C}}^{1 \times \left( {N + 1} \right)}}$ and ${\overline {\bm{\beta}} _k} \triangleq {\left[ {{\bm{\beta}}_k^T\;1} \right]^T}\in {{\mathbb{C}}^{\left( {N + 1} \right) \times 1}}$. For each $\beta _n^k$ and fixing the remaining $N-1$ coefficients, ${\left| {{\mathbf{b}}_k^H{{\overline {\bm{\beta}} }_k}} \right|^2}$ can be expressed as follows:
\begin{align}\label{amplitude2}
{\left| {{\mathbf{b}}_k^H{{\overline {\bm{\beta}}  }_k}} \right|^2} = {C_{k,n}} + {D_{k,n}}\beta _n^k + {E_{k,n}}\sqrt {\beta _n^k},
\end{align}
where ${C_{k,n}} \triangleq {\left| {\sum\nolimits_{l \ne n}^{N + 1} {{{\left[ {{\mathbf{s}}_k^H} \right]}_l}} \sqrt {\beta _l^k} } \right|^2}$, ${D_{k,n}} \triangleq {\left| {{{\left[ {{\mathbf{b}}_k^H} \right]}_n}} \right|^2}$, ${E_{k,n}} \triangleq 2\operatorname{Re} \left\{ {{{\left[ {{\mathbf{b}}_k^H} \right]}_n}{{\left\{ {\sum\nolimits_{l \ne n}^{N + 1} {{{\left[ {{\mathbf{b}}_k^H} \right]}_l}} \sqrt {\beta _l^k} } \right\}}^ * }} \right\}$, and $\beta _{N + 1}^k \triangleq 1, \forall k \in \left\{ {t,r} \right\},n \in {{\mathcal{N}}}$. For given $N-1$ transmission and reflection amplitude coefficients, the design of the $n$th transmission and reflection amplitude coefficients can be formulated as the following problem:
\begin{subequations}\label{P6-X}
\begin{align}
&\mathop {\min }\limits_{{\beta _n^k},{\psi _k}} \;\; \frac{{\overline \gamma _r^{\rm{N}}}}{{{\psi _r}}} + \frac{{\overline \gamma _t^{\rm{N}}\left( {\overline \gamma _r^{\rm{N}} + 1} \right)}}{{{\psi _t}}}\\
\label{C P6-X1}{\rm{s.t.}}\;\;&{C_{k,n}} + {D_{k,n}}\beta _n^k + {E_{k,n}}\sqrt {\beta _n^k}  \ge {\psi _k},\forall k \in \left\{ {t,r} \right\},\\
\label{C P6-X2}&\beta _n^t,\beta _n^r \in \left[ {0,1} \right],\beta _n^t + \beta _n^r = 1,\forall n \in {{\mathcal{N}}}.
\end{align}
\end{subequations}
The potential non-convexity of \eqref{P6-X} lies in the third term of the LHS of \eqref{C P6-X1}, whose convexity depends on the sign of ${E_{k,n}}$. If ${E_{k,n}}\ge 0$, the LHS of \eqref{C P6-X1} is concave with respect to ${\beta _n^k}$. In this case, problem \eqref{P6-X} is convex and can be efficiently solved using CVX~\cite{cvx}. However, if ${E_{k,n}}< 0$, the LHS of \eqref{C P6-X1} is not concave but convex with respect to ${\beta _n^k}$ and problem \eqref{P6-X} is non-convex. To solve this issue, for a given feasible point $\beta _n^{k\left( 0 \right)}$, by employing the first-order Taylor expansion, a lower bound of ${E_{k,n}}\sqrt {\beta _n^k}$ can be obtained as follows:
\begin{align}
{E_{k,n}}\sqrt {\beta _n^k}  \ge \frac{{{E_{k,n}}\left( {\beta _n^k + \beta _n^{k\left( 0 \right)}} \right)}}{{2\sqrt {\beta _n^{k\left( 0 \right)}} }} \triangleq {\Gamma _{k,n}}\left( {\beta _n^k,\beta _n^{k\left( 0 \right)}} \right).
\end{align}
Therefore, for a given feasible point $\beta _n^{k\left( 0 \right)}$ and ${E_{k,n}}< 0$, problem \eqref{P6-X} can be approximated as the following convex optimization problem:
\begin{subequations}\label{P7-X}
\begin{align}
&\mathop {\min }\limits_{{\beta _n^k},{\psi _k}} \;\; \frac{{\overline \gamma _r^{\rm{N}}}}{{{\psi _r}}} + \frac{{\overline \gamma _t^{\rm{N}}\left( {\overline \gamma _r^{\rm{N}} + 1} \right)}}{{{\psi _t}}}\\
\label{C P7-X1}{\rm{s.t.}}\;\;&{C_{k,n}} + {D_{k,n}}\beta _n^k + {\Gamma _{k,n}}\left( {\beta _n^k,\beta _n^{k\left( 0 \right)}} \right) \ge {\psi _k},\forall k \in \left\{ {t,r} \right\},\\
\label{C P7-X2}&\beta _n^t,\beta _n^r \in \left[ {0,1} \right],\beta _n^t + \beta _n^r = 1,\forall n \in {{\mathcal{N}}},
\end{align}
\end{subequations}
which can be solved using CVX~\cite{cvx} to find a suboptimal solution to the original problem \eqref{P6-X}. Similarly, we can alternatingly optimize each $\left\{ {\beta _n^k} \right\}$ with all the other $\left\{ {\beta _l^k} \right\}_{l \ne n}^N$ fixed  to obtain a suboptimal solution to the amplitude coefficient design subproblem.
\subsubsection{Element-wise AO Algorithm} Based on the above two subproblems, an iterative element-wise AO algorithm is developed for solving problem \eqref{P3-X}. In each iteration, the two transmission/reflection phase-shift and amplitude coefficient design subproblems are alternatingly optimized. Within each subproblem, each phase-shift/amplitude coefficient is alternatingly optimized in an element-wise manner. The details of the developed algorithm are summarized in \textbf{Algorithm 1}. The computational complexity of \textbf{Algorithm 1} is analyzed as follows. The main complexity is caused by solving each element-wise phase-shift and amplitude coefficient design problem, which has a complexity of ${\mathcal{O}}\left( 1 \right)$ if the interior point method is employed. As there are two possible phase-shift differences for \eqref{P5-X}, the corresponding complexity for the $N$-element transmission and reflection phase-shift coefficient design problem is ${{\mathcal{O}}}\left(  {2N}  \right)$. The complexity for the $N$-element transmission and reflection amplitude coefficient design problem is ${{\mathcal{O}}}\left( {N}  \right)$. Therefore, the overall computational complexity of \textbf{Algorithm 1} is ${{\mathcal{O}}}\left( {I\left( {3N} \right)} \right)$, where $I$ denotes the number of iterations required for convergence. As a result, the complexity for solving problem \eqref{P2-X} for NOMA and OMA is ${{\mathcal{O}}}\left( {2I\left( {3N} \right)} \right)$ and ${{\mathcal{O}}}\left( {I\left( {3N} \right)} \right)$, respectively, since for NOMA two possible decoding orders have to be considered. Overall, the complexity of the proposed algorithm increases only linearly with $N$, which is of vital importance since the number of STAR elements is usually large in practice.
\begin{algorithm}[!t]\label{method1}
\caption{Element-wise AO algorithm for solving \eqref{P3-X}}
\begin{algorithmic}[1]
\STATE {Randomly initialize feasible $\left\{ {{{\mathbf{q}}_k},k \in \left\{ {t,r} \right\}} \right\}$ and initialize  $\left\{\beta _n^t = \beta _n^r = 0.5,\forall n \in {\mathcal{N}}\right\}$.}
\STATE {\bf repeat}
\STATE \quad Optimize and update the phase-shift coefficients in an element-wise manner by solving \eqref{P5-X}.
\STATE \quad Optimize and update the amplitude coefficients in an element-wise manner by solving \eqref{P6-X} or \eqref{P7-X}.
\STATE {\bf until} the fractional decrease of the objective function value is below a predefined threshold.
\end{algorithmic}
\end{algorithm}
\section{Numerical Results}
\begin{figure}[!b]
    \begin{center}
        \includegraphics[width=2.4in]{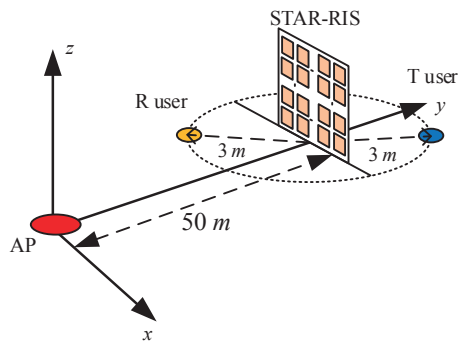}
        \caption{The simulated setup.}
        \label{setup}
    \end{center}
\end{figure}
In this section, numerical results are provided to compare the performance of the proposed algorithm with two benchmark schemes. The considered three-dimensional (3D) simulation setup is shown in Fig. \ref{setup}, where the locations of the BS and the STAR-RIS are $\left( {0,0,0} \right)$ meter and $\left( {0,50,0} \right)$ meter, respectively. The T and R users are randomly located on half-circles centered at the STAR-RIS with a radius of $3$ m. The BS-user channel is modeled as Rayleigh fading channel with path loss exponent of $3.5$, while the BS-STAR-RIS and the STAR-RIS-user channels are modeled as Rician fading channels with path loss exponent of $2.2$ and Rician factor of $3$ dB. The path loss at a reference distance of 1 meter is set to $-30$ dB and the noise power is set to $\sigma ^2 =  - 80$ dBm. The following results were obtained by averaging over 100 channel realizations, where 100 user distributions were first randomly generated, and then the corresponding channel realizations were generated accordingly.\\
\indent For performance comparison, we consider the following two benchmark schemes. (1) \textbf{An upper bound}, where the phase-shift coefficients of the STAR-RIS for transmission and reflection can be independently adjusted~\cite{protocol}. (2) \textbf{Conventional RISs}, where one conventional ${N \mathord{\left/
 {\vphantom {N 2}} \right.
 \kern-\nulldelimiterspace} 2}$-element reflecting-only RIS and one ${N \mathord{\left/
 {\vphantom {N 2}} \right.
 \kern-\nulldelimiterspace} 2}$-element transmitting-only RIS are deployed adjacent to each other. For each benchmark scheme, both NOMA and OMA are considered.\\
\begin{figure}[t!]
\centering
\subfigure[${\overline R _t} = {\overline R _r} = 2$ bit/s/Hz.]{\label{EQoS}
\includegraphics[width= 3in]{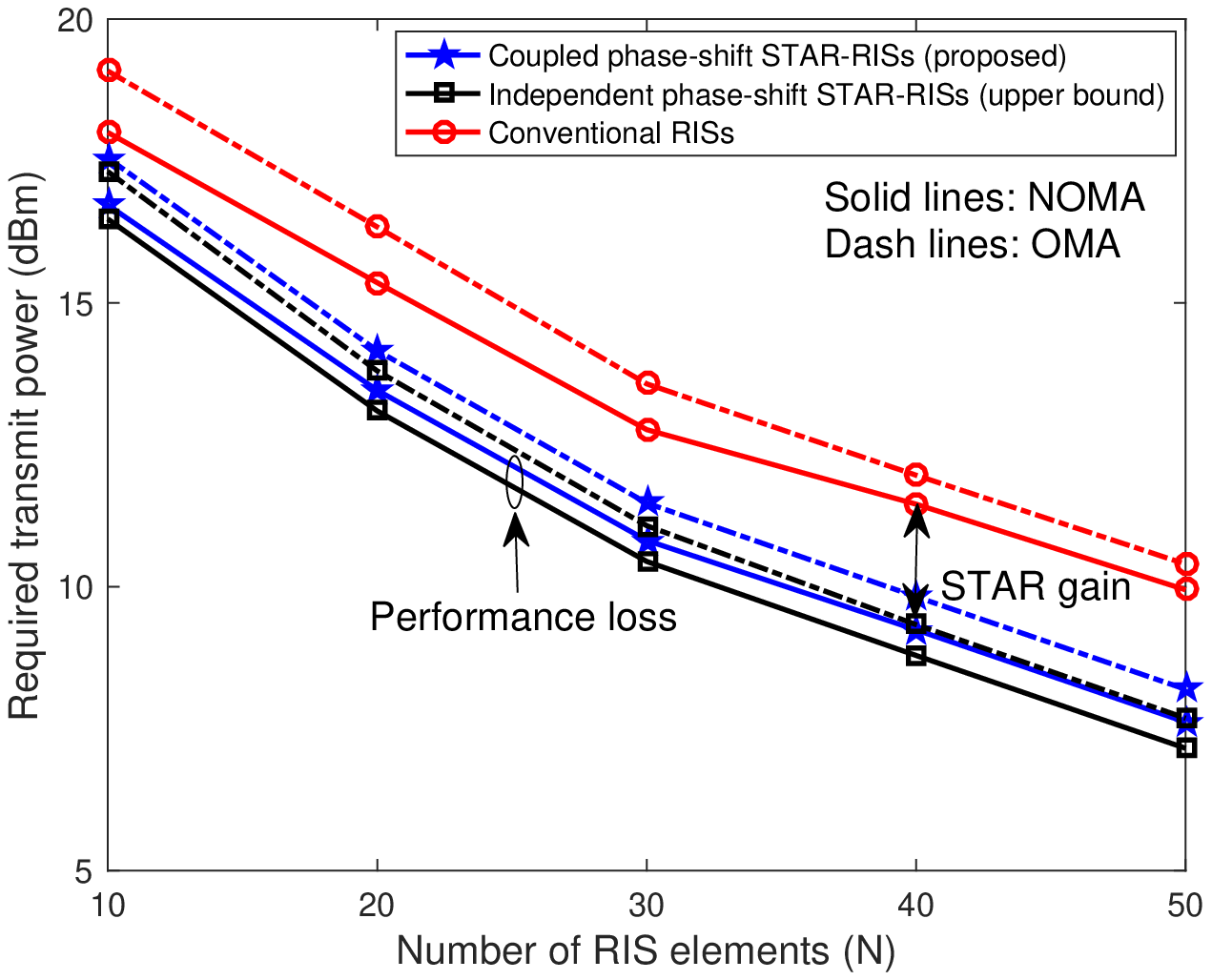}}
\setlength{\abovecaptionskip}{-0cm}
\subfigure[${\overline R _t} = 5$ bit/s/Hz, ${\overline R _r} = 1$ bit/s/Hz.]{\label{HQoS}
\includegraphics[width= 3in]{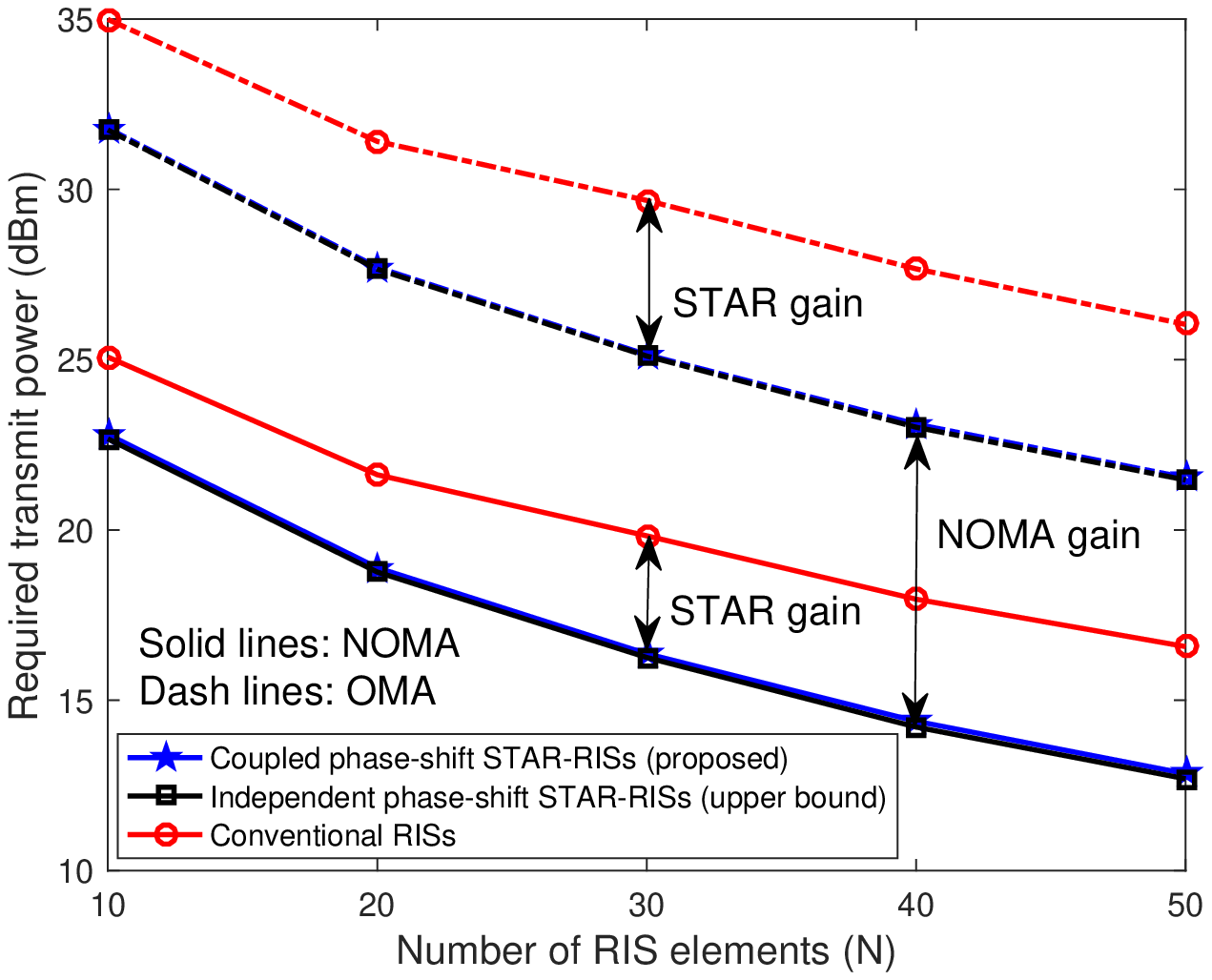}}
\setlength{\abovecaptionskip}{-0cm}
\caption{Power consumption versus $N$.}\label{CvN}
\end{figure}
\indent In Fig. \ref{CvN}, the required power consumption versus the number of RIS elements is studied. We consider two cases, namely symmetric rate requirements (${\overline R _t} = {\overline R _r} = 2$ bit/s/Hz) and asymmetric rate requirements (${\overline R _t} = 5$ bit/s/Hz, ${\overline R _r} = 1$ bit/s/Hz). As can be observed from Fig. \ref{CvN}, for all cases and schemes, the required power consumption decreases as $N$ increases due to the higher transmission/reflection gain. As each element for STAR-RISs has both transmission and reflection coefficients to be optimized (i.e., more degree-of-freedoms (DoFs) to be exploited), STAR-RISs significantly outperform conventional RISs. Regarding the performance loss caused by the coupled phase-shift model, for both NOMA and OMA, a noticeable performance gap can be observed in Fig. \ref{EQoS}, while it becomes negligible in Fig. \ref{HQoS}. The reasons behind this are as follows. For symmetric rate requirements, STAR-RISs tend to maximize the effective channel power gains of both users to minimize the power consumption. Therefore, each element has to be operated in the the real STAR mode (i.e., $\beta _n^t \ne 0$ and $\beta _n^r \ne 0$), where the STAR-RIS employing the independent phase-shift model can exploit more DoFs than that employing the coupled phase-shift model. However, for asymmetric rate requirements, STAR-RISs tend to maximize the effective channel power gain of the user having the higher rate requirement (i.e., the T user) to greatly reduce the power consumption. As a result, each element is optimized to work in the pure transmission mode for serving the T user while the R user is served only thorough the direct link from the BS. In this case, the effect of the coupled phase-shift model can be ignored. Moreover, it can be observed that the performance gain of NOMA over OMA is more pronounced for the case of asymmetric rate requirements compared to the case of symmetric rate requirements. This is expected since as explained above, STAR-RISs prefer to only enhance the channel power gain of the rate-hungry user for the case of asymmetric rate requirements, thus yielding a significant channel quality difference where NOMA achieves a higher performance gain.
\section{Conclusions}
A STAR-RIS aided communication system was investigated under a realistic coupled phase-shift model. Considering both NOMA and OMA, the transmission and reflection coefficient optimization problem was formulated for minimization of the required power consumption of the AP subject to user rate requirements. The formulated non-convex problem was efficiently solved with an element-wise AO algorithm, whose complexity scales only linearly with the number of STAR elements. Numerical results showed that STAR-RISs outperform conventional reflecting/transmitting-only RISs. For STAR-RISs, the performance degradation caused by the coupled phase-shift model is noticeable for symmetric user rate requirements, while it is negligible for asymmetric user rate requirements. Moreover, the performance gain of NOMA over OMA was shown to be significant for asymmetric user rate requirements.
\vspace{-0.2cm}
\bibliographystyle{IEEEtran}
\bibliography{mybib}
 \end{document}